\documentclass[a4paper,fleqn,usenatbib]{mnras}

\usepackage[T1]{fontenc}
\usepackage{ae,aecompl}
\usepackage{color}
\usepackage{graphicx}	
\usepackage{amsmath}	
\usepackage{amssymb}	

\newcommand {\eV}       {\,\rm eV}
\newcommand {\pc}       {\,\rm pc}
\newcommand {\kpc}      {\,\rm kpc}
\newcommand {\Msun}     {\, M_{\sun}}
\newcommand {\kms}      {\,\rm km\,s^{-1}}
\newcommand {\psiDM}    {\psi {\rm DM}}
\newcommand {\sref}[1]  {Section~\ref{#1}}
\newcommand {\fref}[1]  {Figure~\ref{#1}}
\newcommand {\tref}[1]  {Table~\ref{#1}}
\newcommand {\eref}[1]  {Equation~(\ref{#1})}

\title[Determining $\psiDM$ particle mass from dSphs]{Jeans Analysis for Dwarf Spheroidal Galaxies in Wave Dark Matter}

\author[S-R. Chen et al.]{
Shu-Rong Chen,$^{1}$\thanks{E-mail: f97222062@ntu.edu.tw (SRC)}
Hsi-Yu Schive,$^{2}$
Tzihong Chiueh$^{1,3,4}$
\\
$^{1}$Department of Physics, National Taiwan University, 10617 Taipei, Taiwan\\
$^{2}$National Center for Supercomputing Applications, Urbana, IL, 61801, USA\\
$^{3}$Institute of Astrophysics, National Taiwan University, 10617 Taipei, Taiwan\\
$^{4}$Center for Theoretical Sciences, National Taiwan University, 10617 Taipei, Taiwan
}

\date{Accepted XXX. Received YYY; in original form ZZZ}

\pubyear{2016}

\begin{document}
\label{firstpage}
\pagerange{\pageref{firstpage}--\pageref{lastpage}}
\maketitle

\begin{abstract}
Although still under debate, observations generally suggest that dwarf spheroidal (dSph) galaxies exhibit large
constant-density cores in the centers, which can hardly be explained by
dissipationless cold dark matter simulations without baryonic feedback. Wave dark matter ($\psiDM$),
characterized by a single parameter, the dark matter particle mass $m_{\psi}$,
predicts a central soliton core in every galaxy arising from quantum pressure
against gravity. Here we apply Jeans analysis assuming a soliton core profile to the kinematic data of
eight classical dSphs so as to constrain $m_{\psi}$, and obtain 
$m_{\psi}=1.18_{-0.24}^{+0.28}\times10^{-22}\eV$ and 
$m_{\psi}=1.79_{-0.33}^{+0.35}\times10^{-22}\eV~(2\sigma)$ using the
observational data sets of \citet{Walker2007} and \citet{Walker2009b},
respectively. We show that the estimate of $m_{\psi}$ is sensitive
to the dSphs kinematic data sets and is robust to various models of
stellar density profile.
We also consider multiple stellar subpopulations in dSphs and find consistent results. This mass range
of $m_{\psi}$ is in good agreement with other independent estimates, such as
the high-redshift luminosity functions, the reionization history, and the
Thomson optical depth to the cosmic microwave background. 
\end{abstract}


\section{Introduction}
\label{sec:Intro}

Dark matter, which constitutes $\sim 26\%$ of the total energy density
\citep{Planck2015}, plays an important role in the structure formation
of the Universe. The conventional cold dark matter (CDM) model has been
successful at explaining observations on large scales, such as cosmic microwave
background (CMB) and large-scale galaxy structure resulting from primordial
fluctuations \citep{Tegmark2006, Cole2005, Frenk&White2012, Bennett2013}.
However, several discrepancies between observations and dissipationless CDM
simulations on the galactic scale still persist. There are three major problems.
(i) ``missing satellite problem'': CDM simulations predict an order of
magnitude more subhalos than observed satellite galaxies in the
Milky Way \citep{Klypin1999, Moore1999}. 
(ii) ``too-big-to-fail problem'': the average central density of the most
massive CDM subhalos are significantly higher than that of the most luminous
dwarf spheroidal (dSph) galaxies derived from their kinematic data
\citep{Boylan2012, Boylan2011, Tollerud2012}. 
(iii) ``cusp-core problem'': CDM simulations predict a universal
Navarro-Frenk-White (NFW) halo density profile
\citep{Navarro1997} with a cuspy
central density, while observations of dSphs
\citep{Moore1994, Flores1994, Kleyna2002, Goerdt2006,Walker2011, Amorisco2013}
and low surface brightness galaxies
\citep{deBlok2001, deBlok2002, deBlok2005, Burkert1995, Borriello2001, Oh2008}
are, in general, better described by a cored density profile.

Generally speaking, there are two categories of solutions to these small-scale
issues. The first class of approaches considers additional baryonic physics
in CDM simulations, such as
feedback from supernova explosions and stellar wind
\citep{Navarro1996, Read2005, Governato2010, Governato2012,
Mashchenko2008, Pontzen2014}, energy transfer from
dynamical friction of compact baryonic objects \citep{El-Zant2001},
ram pressure stripping \citep{Arraki2014},
tidal stripping \citep{Brooks2013}, suppression of star formation rate by 
photoionization due to early reionization
\citep{Bullock2000, Benson2002, Somerville2002},
blazer \citep{Pfrommer2012}, and
cosmic-ray heating \citep{Wadepuhl2011}. See \citet{Weinberg2013} for
a comprehensive review.
The second class of approaches adopts alternative dark matter models,
such as warm dark matter \citep{Colin2000, Maccio2012, Lovell2012}, 
self-interacting dark matter \citep{Spergel2000, Vogelsberger2012},
and axion/scalar field dark matter
\citep{Turner1983, Khlopov1985, Sin1994, Guzman2000, Peebles2000, Goodman2000, Hu2000,
Matos2000,Sahni2000, Bohmer2007, Sikivie2009, Chavanis2011,Robles2013,Guth2015,Davidson2015},
which naturally produce constant-density cores and a cut off in the matter
power spectrum.

In this work, we focus on scalar field dark matter composed of extremely
light bosons with negligible self-interaction, known as \emph{$\psiDM$}
\citep{Schive2014a, Marsh&Silk2013} or \emph{fuzzy} dark matter
\citep{Hu2000}. In this scenario, dark matter consists of axion-like particles
proposed by string theory \citep{Arvanitaki2010, Svrcek2006} or non-QCD
axions \citep{Chiueh2014}. By assuming a dark matter particle mass of
$m_{\psi} \sim 10^{-22}\eV$, the de Broglie wave nature becomes manifest on
astrophysical scale. Here uncertainty principle counters gravity below a
Jeans scale, simultaneously leading to kpc-scale central density cores
in dSph-sized subhalos and suppressing the abundance of subhalos with
masses below $\sim 10^{10}\Msun$
\citep{Hu2000, Marsh2010, Schive2014a, Schive2014b}. It thus provides a
plausible solution to the small-scale issues in CDM. Moreover,
on large scales $\psiDM$ is statistically indistinguishable from CDM
\citep{Woo2009, Schive2014a, Marsh&Silk2013, Bozek2015, Widrow1993}. 
For these reasons, $\psiDM$ has become a viable candidate for dark matter. 

One of the key features in $\psiDM$ is that it has only one free parameter,
$m_{\psi}$, the dark matter particle mass. It is thus crucial to validate
whether the estimates of $m_{\psi}$ from various independent observational
constraints are consistent with each other.
Using CMB and galaxy clustering data, \citet{Hlozek2015} obtained 
$m_{\psi}>10^{-24}\eV$. \citet{Schive2016} used high-redshift galaxy luminosity
function to derive $m_{\psi}>1.2\times10^{-22}\eV$
(see also \citet{Bozek2015,Corasaniti2016}). \citet{Sarkar2016} used damped Lyman-$\alpha$
observations and found $m_{\psi}>10^{-23}\eV$. \citet{Marsh&Pop2015}
explored multiple stellar subpopulations in Fornax and Sculptor
and estimated $m_{\psi}<1.1\times 10^{-22}\eV$. \citet{Lora2012, Lora2014} used
the longevity of the cold clumps in Ursa Minor and 
Sextans and found $m_{\psi} \sim 0.3 \-- 1\times10^{-22}\eV$ and
$m_{\psi} \sim 0.12 \-- 8\times10^{-22}\eV$, respectively.
Using newly discovered ultra-faint dSphs, \citet{Calabrese2016} estimated
$m_{\psi} \sim 3.7 \-- 5.6\times10^{-22}\eV$. 
According to these studies, which cover a variety of astrophysical probes,
the viable $\psiDM$ particle mass generally lies in the range
$10^{-23} \-- 10^{-21}\eV$. See \citet{Marsh2015a} for a comprehensive review.

DSph galaxies are the most dark-matter-dominated systems with
mass-to-light ratios exceeding $>100$ \citep{Mateo1998,Kleyna2005}
and with little gas and no recent star formation
\citep{Smecker-Hane1994, Tolstoy2003, Venn2004}.
Large spectroscopic surveys provide rich stellar kinematics and metallicity
data of dSphs, allowing detailed studies on dark matter properties
from dynamical modeling. We refer to \citet{Battaglia2013} for a
comprehensive review on this subject.
Several works suggest that the dark matter density
profiles are cored rather than cuspy \citep{Kleyna2003,Goerdt2006,Sanchez-Salcedo2006,
 Battaglia2008, Jardel2012, Walker2011, Amorisco&Evans2012a}, 
although debate still remains \citep{Strigari2010, Strigari2014, Jardel2013}. 
In this work, we apply Jeans analysis to the kinematic data of eight classical dSphs to determine
$m_{\psi}$. Most importantly, we investigate (i) whether the estimates
of $m_{\psi}$ from different dSphs are in good agreement with each other,
and (ii) whether the combined constraint of $m_{\psi}$ from the eight classical dSphs
is consistent with other independent observational constraints described
previously.

The paper is structured as follows. We describe the procedure of our Jeans
analysis in \sref{sec:Jeans} and show results in \sref{sec:Results}.
In \sref{sec:Discussion}, we address various uncertainties in the analysis
and discuss some extended models. Finally, we summarize our findings in
\sref{sec:Conclusion}.

\section{Jeans Analysis}
\label{sec:Jeans}
In order to constrain the dark matter particle mass $m_{\psi}$ by the stellar
kinematics of dSphs, we regard stars as tracers of the gravitational potential
dominated by the dark matter and assume the tidal disturbance is negligible.
Dynamical equilibrium of stars in dSphs is
supported by velocity dispersion with negligible rotation \citep{Mateo1998}.
Assuming spherical symmetry, the Jeans equation
\citep{GalacticDynamics2008} relates the stellar phase-space distribution to
the dark matter halo as
\begin{equation}
\frac{1}{\nu}\frac{d}{dr}(\nu \bar{v_r^2})+2\frac{\beta\bar{v_r^2}}{r}=-\frac{GM(r)}{r^2},
\label{eq:jeans}
\end{equation}
where $\nu(r)$, $\bar{v_r}(r)$, $\bar{v_t}(r)$, and $\beta(r)\equiv 1-\bar{v_t^2}/2\bar{v_r^2}$
are the stellar number density, radial velocity dispersion, tangential velocity dispersion, and orbital anisotropy,
respectively. $M(r)$ is the enclosed mass of the dark matter. By assuming
$\beta = \rm{constant}$, \eref{eq:jeans} has the solution \citet{Lokas&Mamon2003}:
\begin{equation}
\nu\bar{v^2_r}=Gr^{-2\beta}\int_r^{\infty}s^{2\beta-2}\nu(s)M(s)ds.
\label{eq:jeanssolution}
\end{equation}
By projecting \eref{eq:jeanssolution} along the line of sight, we get \citep{GalacticDynamics2008}
\begin{equation}
\sigma_p^2(R)=\frac{2}{\Sigma(R)}\int_{R}^{\infty}\biggl (1-\beta\frac{R^2}{r^2}\biggr ) \
              \frac{\nu \bar{v_r^2}r}{\sqrt{r^2-R^2}}dr,
\label{eq:jeansproject}
\end{equation}
where $R$, $\Sigma(R)$, and $\sigma_p(R)$ are the projected radius, stellar
surface density, and line-of-sight velocity dispersion, respectively.

In the following, we address in detail (1) the dark matter mass profile in the $\psiDM$
model, (2) the stellar density and velocity dispersion profiles in the Jeans
analysis, and (3) the Markov chain Monte Carlo algorithm for constraining the
posterior distribution of $m_{\psi}$.

\subsection{Wave Dark Matter Halo}
\label{subsec:wave dark matter model}
From numerical simulations, \citet{Schive2014a} found a gravitationally
self-bound, constant-density core at the center of each $\psiDM$ halo, which
connects to an NFW profile at a larger radius. This cored profile satisfies a
soliton solution and can be well fitted by \citep{Schive2014a}
\begin{equation}
\rho_{\rm{soliton}}(r) = \frac{1.9~(m_\psi/10^{-23}\eV)^{-2}(r_c/\pc)^{-4}}
                         {[1+9.1\times10^{-2}(r/r_c)^2]^8}~10^{12}\Msun\pc^{-3},
\label{eq:SolitonFit}
\end{equation}
which has two free parameters, $m_{\psi}$ and $r_c$,
where $r_c$ is the core radius defined as the radius at which the density drops
to one-half its peak value.
The main goal of this work is thus to determine both $r_c$ and $m_{\psi}$ for each dSph.
Note that the mass profile, $M(r)= 4\pi\int_{0}^{r}s^2\rho(s)ds$,
can be calculated analytically from \eref{eq:SolitonFit}.
See Appendix \ref{sec:SolitonMassProfile} for the explicit form.

\citet{Schive2014a} shows that, in general, the transition radius between the
inner soliton and the outer NFW halo is greater that $3\,r_c$, which is a few
times greater than the observed half-light radii of dSphs. It is therefore
reasonable to assume that all stars reside within the central soliton
and ignore the outer NFW halo in the first place when conducting the Jeans
analysis. For dSphs with stars possibly extending beyond $3\,r_c$ (e.g., Fornax),
we also extend the main analysis to include outer NFW halos
(see \sref{subsec:SolitonNFW}).

\subsection{Stellar Density and Velocity Dispersion}
\label{subsec:stellar density and velocity dispersion}
Stellar surface density of dSphs are commonly fitted by King, Sersic, or
Plummer models \citep[e.g.,][]{Irwin1995}. Following \citet{Walker2009b}, we
first adopt a Plummer profile, $I(R)=L(\pi R_h^2)^{-1}[1+R^2/R_h^2]^{-2}$,
where $L$ is the total luminosity and $R_h$ is the radius enclosing $0.5\,L$.
We use $R_h$ derived from \citet{Walker2009b}.
Further assuming a constant mass-to-light ratio, the three-dimensional
density profile can be derived from the surface density profile by the
Abel transform as
$\nu(r)=3L(4\pi R_h^3)^{-1}[1+r^2/R_h^2]^{-5/2}$.
We also investigate other stellar density models in
\sref{subsubsec:StellarDens} so as to consolidate the results.

For the stellar velocity dispersion, we first adopt the data of eight classical dSphs in
\citet{Walker2007} and \citet{Walker2009b, Walker2013}
(see Figs. \ref{fig:vel_disp} and \ref{fig:vel_disp_w07}), where the error bars indicate $1\sigma$
uncertainty. We shall also discuss the results using different observational data sets
in \sref{subsubsec:VelDisp}.

\subsection{Markov Chain Monte Carlo}
\label{subsec:MCMC}
To constrain the range of $m_{\psi}$, we follow \citet{Walker2009b,Salucci2012} and
compare the empirical line-of-sight velocity dispersion, $\sigma_{V_0}(R_i)$,
with the projected velocity dispersion, $\sigma_p(R_i)$, estimated from our
Jeans analysis using \eref{eq:jeansproject}.
We explore a three-dimensional parameter vector set
$\vec{\theta} \equiv \left \{ -\log_{10}(1-\beta),\,\log_{10}[r_c/\rm{pc}],\,
\log_{10}[m_\psi /10^{-23}\rm{eV}] \right \}$
with uniform priors over the ranges
$-1\leq-\log_{10}(1-\beta)\leq1$,
$1\leq \log_{10}[r_c/\rm{pc}]\leq4$, and
$0\leq \log_{10}[m_\psi /10^{-23}\rm{eV}]\leq3$.
We take the likelihood function
\begin{equation}
L(\vec{\theta}) = \prod_{i=1}^N \frac{1}{\sqrt{2\pi\rm{Var}[\sigma_{V_0}(R_i)]}}
\exp \biggl [ -\frac{1}{2}\frac{(\sigma_{V_0}(R_i)-\sigma_p(R_i))^2}{\rm{Var}[\sigma_{V_0} (R_i)]} \biggr ],
\label{eq:likelihood}
\end{equation}
where $\rm{Var}[\sigma_{V_0}(R_i)]$ is the square of the observational
uncertainty associated with $\sigma_{V_0}(R_i)$ and $N$ is the number of bins
in the velocity dispersion profile. We get a
posterior probability distribution of each model parameter using a
Markov chain Monte Carlo (MCMC) analysis with the Metropolis-Hastings algorithm
\citep{Metropolis1953, Hastings1970}. We use the MCMC engine \emph{CosmoMC}
\citep{Lewis2002ah,Lewis1999bs} as a generic sampler and run four parallel chains.
The code computes the R-statistic of Gelman and Rubin \citep{An98stephenbrooks}
as the convergence criterion and stops iterations when this value is less than
1.01 for each model parameter. The first $30\%$ accepted steps are discarded
as burn-in.

To account for the observational uncertainty of $R_h$, we follow
\citet{Salucci2012} and let $R_h$ sample randomly from a Gaussian
distribution with mean and standard deviation equal to the observations
\citep{Irwin1995}. We also investigate the case with a fixed $R_h$ in
\sref{subsubsec:StellarDens}.

\section{Results}
\label{sec:Results}

\begin{figure}
\centering
\includegraphics[width=\columnwidth]{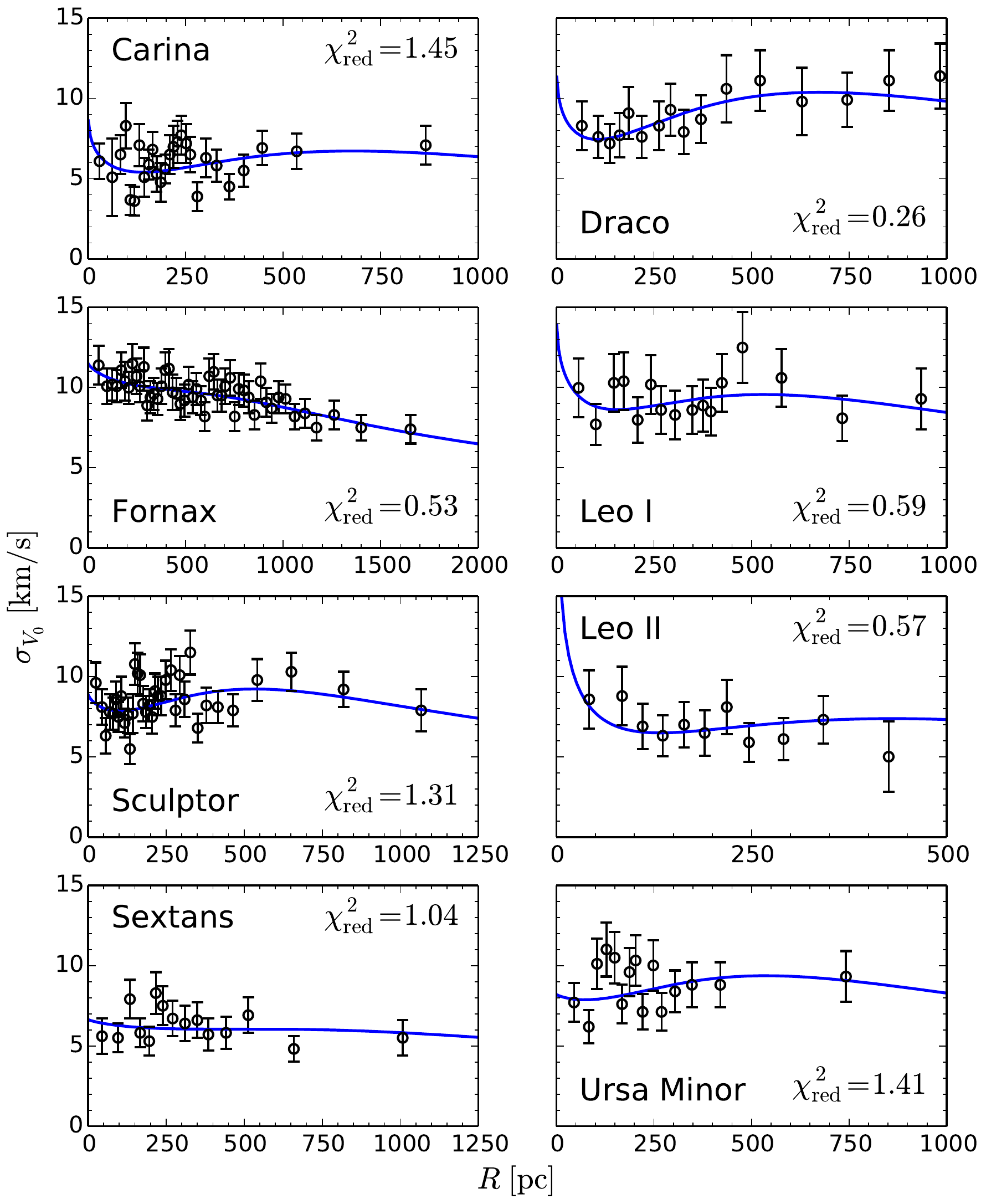}
\caption{
Projected velocity dispersion profiles for the eight classical dSphs in the
observational data set of \citet{Walker2009b}.
Error bars show the empirical profiles with $1\sigma$ uncertainties.
Solid lines show the best-fit profiles obtained in our Jeans analysis using
the soliton density profile in $\psiDM$ (\eref{eq:SolitonFit}), which provide
satisfactory fits to the observations. The reduced chi-square, $\chi_{\rm{red}}^2$,
of each dSph is also shown. The confidence intervals of the model parameters
for each dSph are listed in \tref{table:fit}.
}
\label{fig:vel_disp}
\end{figure}

We apply the Jeans analysis described above to the observational data
of eight classical dSphs.
\fref{fig:vel_disp} shows the empirical velocity dispersions
of \citet[][hereafter W09]{Walker2009b} and the
estimated best-fit dispersions, which correspond to the maximum likelihood points
in our MCMC chains with $m_{\psi}$ confined in the $2\sigma$ range of the
combined constraint from eight dSphs. Clearly, the soliton core profile 
in the $\psiDM$ model provides satisfactory fits to the observations. 
The reduced chi-square, $\chi_{\rm{red}}^2$, for each dSph assuming 
three free parameters lies in the range $0.26 \-- 1.45$.

\begin{figure}
\centering
\includegraphics[width=\columnwidth]{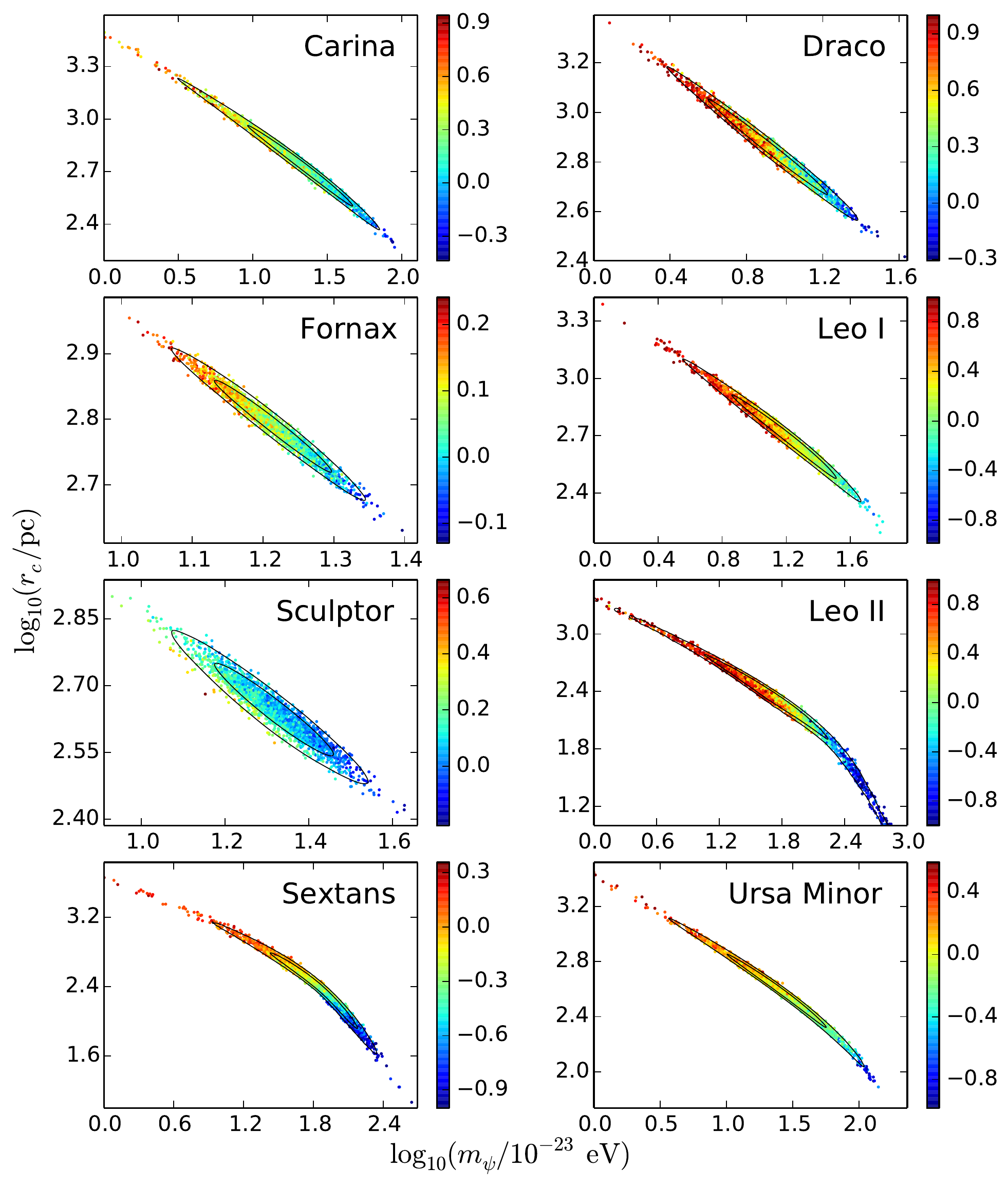}
\caption{
Posterior distributions of $m_{\psi}$ and $r_{c}$ colored by $\beta$ for each dSph in our MCMC
analysis. Contours show the $1\sigma$ and $2\sigma$ confidence regions.
The confidence intervals of the model parameters for each dSph are also
listed in \tref{table:fit}.
}
\label{fig:mb_rc_contour_w09}
\end{figure}

\fref{fig:mb_rc_contour_w09} shows the $1\sigma\,(68\%)$ and $2\sigma\,(95\%)$
contours for the posterior distributions of $m_{\psi}$ and $r_{c}$.
The corresponding values of $\beta$ are shown in the colored scatter plot.
Notice that there is a clear correlation between $m_{\psi}$ and $r_c$.
It results from the fact that a significant fraction of stars are located
within the soliton core radius, where the dark matter density is roughly a
constant and thus $m_{\psi}$ and $r_c$ become degenerate as
$m_{\psi} \propto r_c^{-2}$ (see \eref{eq:SolitonFit}). It therefore relies
on stars outside $r_c$ to break this degeneracy.
The mean correlation is found to be $m_{\psi} \propto r_c^{-1.4}$,
shallower than the fully degenerate case. It explains the tendency that
the tangential anisotropy increases with decreasing $r_c$ and increasing
$m_{\psi}$. By inserting $m_{\psi} \propto r_c^{-1.4}$ into \eref{eq:SolitonFit},
we have $\rho_{\rm{soliton}} \propto r_c^{-1.2}$. Therefore, a smaller $r_c$
corresponds to a higher core density, which needs a larger amount of tangential
anisotropy to counter gravity and match the observed velocity dispersion profile.

\tref{table:fit} lists the means, $1\sigma$, and $2\sigma$ confidence intervals of
$r_c$, $m_{\psi}$, and $\beta$ for each dSph using the one-dimensional posterior
distribution obtained from MCMC chains.
We estimate the combined constraint by multiplying the one-dimensional posterior
distributions of all dSphs based on the assumption that each dSph gives independent
constraint on $m_{\psi}$.
It leads to $1\sigma\,(2\sigma)$ confidence intervals of
$m_{\psi}=1.79_{-0.17(-0.33)}^{+0.17(+0.35)}\times10^{-22}\eV$ and a
corresponding reduced chi-square of particle mass of $\chi_{\rm{red}}^{2}=1.2$.
We also estimate the combined constraint by fitting all eight dSphs simultaneously,
and the results are very consistent with the values given above.
See Appendix \ref{sec:Joint analysis} for details.

Note that the combined constraint is dominated by Fornax since it has the
smallest variance among the eight dSphs. Also note that the anisotropy
parameters $\beta$ of some dSphs have only $1\sigma$ constraints. We have
adopted a larger range for the uniform prior of $\beta$ and also a different
anisotropy prior distribution (see \sref{subsubsec:Anisotropy}) 
and verified that the estimates of $m_{\psi}$ are consistent with \tref{table:fit}.
\begin{table*}
\begin{center}
\caption{Constraints on model parameters from MCMC analysis.
Values represent the means and $1\sigma$ ($2\sigma$) confidence intervals. The first and second rows
for each dSph represent the results estimated from the kinematic data sets of W09 and W07, respectively.
The anisotropy parameters of some dSphs have only $1\sigma$ constraints.}
\label{table:fit}
\begin{tabular}{llrr}
\hline
Galaxy     & $\log_{10}[m_\psi /10^{-23}\rm{eV}]$ & $\log_{10}[r_c/\rm{pc}]$             & $-\log_{10}(1-\beta)$                            \\
\hline                                                                                                                                      
Carina     & $1.29^{+0.29(+0.50)}_{-0.18(-0.56)}$ & $2.76^{+0.13(+0.35)}_{-0.18(-0.33)}$ & $\hphantom{-}0.24^{+0.15(+0.35)}_{-0.18(-0.33)}$ \\
           & $1.09^{+0.18(+0.32)}_{-0.14(-0.35)}$ & $2.87^{+0.09(+0.22)}_{-0.11(-0.21)}$ & $\hphantom{-}0.29^{+0.10(+0.22)}_{-0.11(-0.20)}$ \\
Draco      & $0.91^{+0.20(+0.40)}_{-0.20(-0.40)}$ & $2.86^{+0.12(+0.25)}_{-0.12(-0.24)}$ & $\hphantom{-}0.53^{+0.35\hphantom{(+0.00)}}_{-0.24\hphantom{(-0.00)}}$ \\
           & $1.05^{+0.17(+0.31)}_{-0.15(-0.34)}$ & $2.75^{+0.10(+0.21)}_{-0.10(-0.20)}$ & $\hphantom{-}0.16^{+0.13(+0.31)}_{-0.16(-0.29)}$ \\
Fornax     & $1.21^{+0.06(+0.10)}_{-0.05(-0.11)}$ & $2.79^{+0.05(+0.10)}_{-0.05(-0.09)}$ & $\hphantom{-}0.07^{+0.06(+0.11)}_{-0.06(-0.11)}$ \\
           & $0.99^{+0.07(+0.13)}_{-0.07(-0.15)}$ & $2.90^{+0.05(+0.11)}_{-0.05(-0.10)}$ & $\hphantom{-}0.07^{+0.06(+0.12)}_{-0.06(-0.12)}$ \\
Leo I      & $1.18^{+0.25(+0.42)}_{-0.18(-0.46)}$ & $2.71^{+0.13(+0.31)}_{-0.16(-0.29)}$ & $\hphantom{-}0.33^{+0.22(+0.57)}_{-0.32(-0.48)}$ \\
           & $1.08^{+0.27(+0.44)}_{-0.17(-0.49)}$ & $2.78^{+0.12(+0.31)}_{-0.17(-0.29)}$ & $\hphantom{-}0.41^{+0.18(+0.50)}_{-0.28(-0.33)}$ \\
Leo II     & $1.71^{+0.39(+1.03)}_{-0.42(-0.85)}$ & $2.30^{+0.37(+0.71)}_{-0.24(-0.94)}$ & $\ge 0.21^{\hphantom{+0.00}\hphantom{(+0.00)}}_{\hphantom{+0.00}\hphantom{(+0.00)}}$\\
           & $1.61^{+0.30(+0.64)}_{-0.29(-0.70)}$ & $2.45^{+0.20(+0.45)}_{-0.21(-0.48)}$ & $\hphantom{-}0.40^{+0.46\hphantom{(+0.00)}}_{-0.28\hphantom{(+0.00)}}$ \\
Sculptor   & $1.31^{+0.10(+0.18)}_{-0.08(-0.19)}$ & $2.65^{+0.06(+0.14)}_{-0.07(-0.13)}$ & $\hphantom{-}0.09^{+0.08(+0.18)}_{-0.09(-0.17)}$ \\
           & $1.11^{+0.14(+0.26)}_{-0.11(-0.26)}$ & $2.77^{+0.08(+0.17)}_{-0.09(-0.17)}$ & $\hphantom{-}0.22^{+0.10(+0.27)}_{-0.14(-0.25)}$ \\
Sextans    & $1.79^{+0.33(+0.53)}_{-0.19(-0.58)}$ & $2.41^{+0.31(+0.61)}_{-0.30(-0.64)}$ &            $-0.31^{+0.39(+0.46)}_{-0.19(-0.63)}$ \\
           & $1.33^{+0.22(+0.38)}_{-0.13(-0.44)}$ & $2.78^{+0.12(+0.35)}_{-0.17(-0.32)}$ &            $-0.12^{+0.13(+0.25)}_{-0.12(-0.26)}$ \\
Ursa Minor & $1.39^{+0.29(+0.55)}_{-0.24(-0.65)}$ & $2.59^{+0.18(+0.42)}_{-0.19(-0.44)}$ &            $-0.01^{+0.19(+0.39)}_{-0.14(-0.40)}$ \\
\hline
\end{tabular}
\end{center}
\centering
\end{table*}

\begin{table*}
\begin{center}
\caption{Summary of various stellar models applied to individual dSphs.
The results generally agree with \tref{table:fit} and reveal insensitivity to these variations.
}
\label{table:discussion}
\begin{tabular}{llll}
\hline
Galaxy   & $\log_{10}[m_\psi /10^{-23}\rm{eV}]$ & $\log_{10}[r_c/\rm{pc}]$ & Description                                    \\ \hline
Fornax   & $1.09^{+0.07}_{-0.06}$               & $2.85^{+0.05}_{-0.06}$   & velocity dispersion (Amorisco and Evans 2012b) \\
         & $1.15^{+0.07}_{-0.06}$               & $2.84^{+0.05}_{-0.05}$   & stellar density (generalized Plummer)          \\
         & $1.23^{+0.05}_{-0.07}$               & $2.78^{+0.06}_{-0.05}$   & soliton + NFW model                            \\
         & $0.91^{+0.26}_{-0.09}$               & $2.96^{+0.07}_{-0.19}$   & stellar subpopulations                         \\
Leo I    & $1.19^{+0.32}_{-0.20}$               & $2.70^{+0.14}_{-0.20}$   & velocity dispersion (tidal effect)             \\
         & $0.92^{+0.42}_{-0.21}$               & $2.86^{+0.13}_{-0.24}$   & stellar density (generalized Plummer)          \\
Draco    & $0.85^{+0.34}_{-0.19}$               & $2.90^{+0.12}_{-0.19}$   & stellar density (generalized Plummer)          \\
Sextans  & $0.51^{+0.23}_{-0.44}$               & $3.31^{+0.25}_{-0.14}$   & velocity dispersion (VLT)                      \\
         & $1.64^{+0.48}_{-0.14}$               & $2.55^{+0.24}_{-0.44}$   & stellar density (exponential profile)          \\
Carina   & $1.41^{+0.44}_{-0.18}$               & $2.67^{+0.14}_{-0.28}$   & velocity dispersion (tidal effect)             \\
Sculptor & $1.08^{+0.28}_{-0.22}$               & $2.78^{+0.14}_{-0.14}$   & stellar subpopulations (OM anisotropy)         \\
         & $1.21^{+0.17}_{-0.15}$               & $2.70^{+0.10}_{-0.11}$   & stellar subpopulations (constant anisotropy)   \\ \hline
\end{tabular}
\end{center}
\end{table*}

\begin{figure}
\centering
\includegraphics[width=\columnwidth]{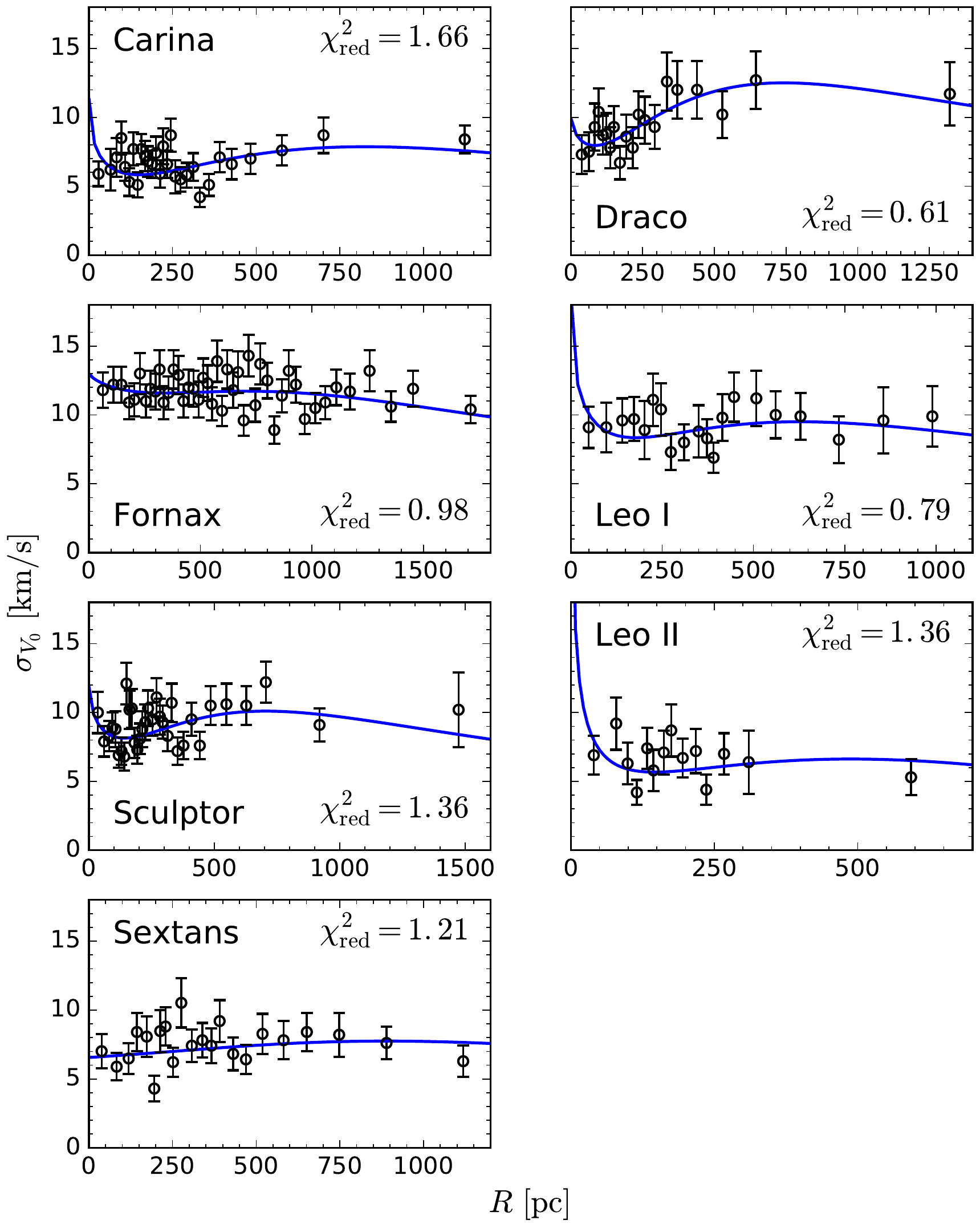}
\caption{
Same as \fref{fig:vel_disp} but for the observational data set of
\citet{Walker2007}. The confidence intervals of the model parameters for each
dSph are listed in \tref{table:fit}.
}
\label{fig:vel_disp_w07}
\end{figure}

\begin{figure}
\centering
\includegraphics[width=\columnwidth]{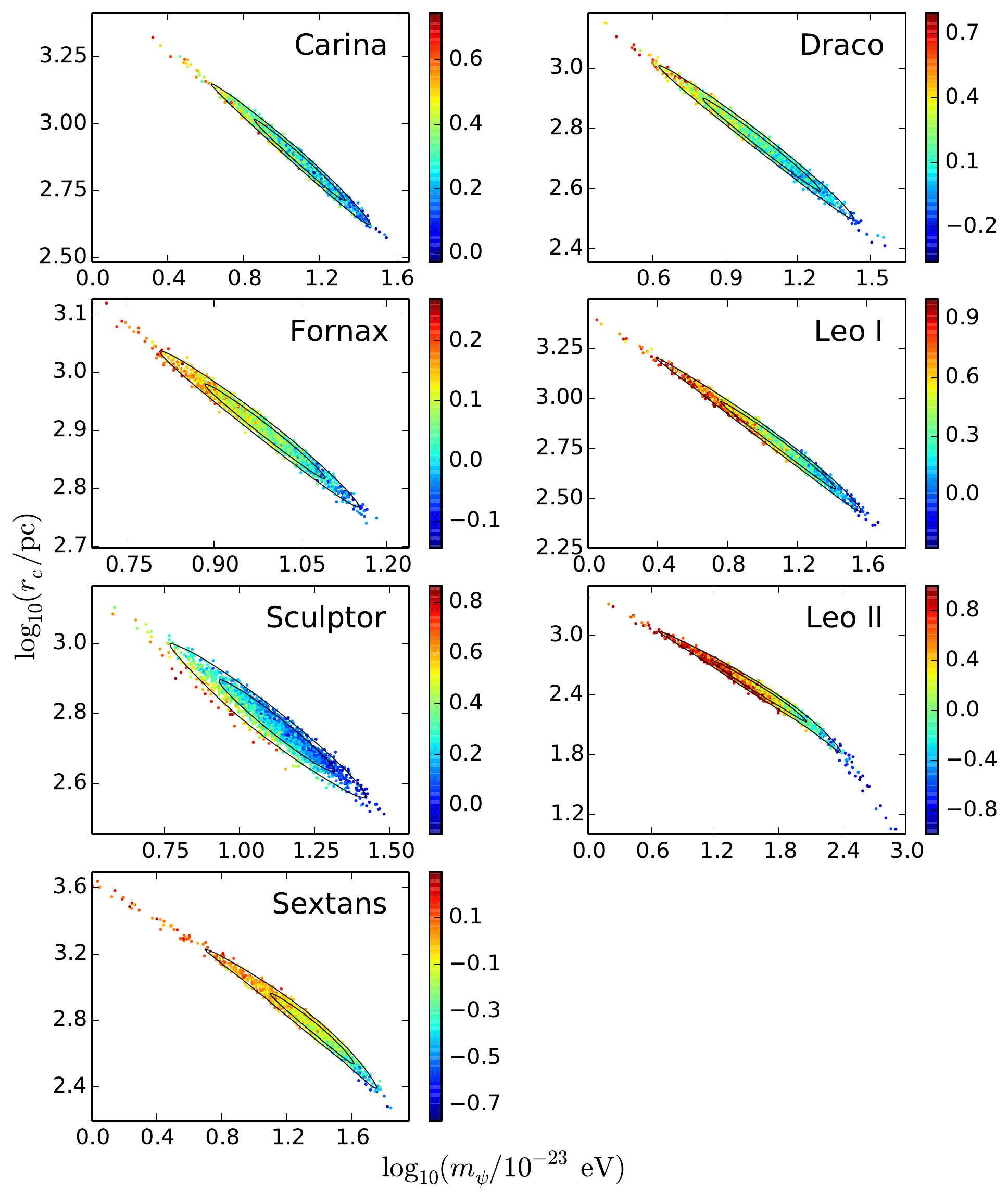}
\caption{
Same as \fref{fig:mb_rc_contour_w09} but for the observational data set of
\citet{Walker2007}. The confidence intervals of the model parameters for each
dSph are listed in \tref{table:fit}.
}
\label{fig:mb_rc_contour_w07}
\end{figure}

To further ascertain whether our estimate of $m_{\psi}$ is sensitive to the
adopted observational data sets, we also apply the same Jeans analysis to the data of seven
dSphs in \citet[][hereafter W07]{Walker2007}, which does not include Ursa Minor.
The results are shown in Figures \ref{fig:vel_disp_w07} and \ref{fig:mb_rc_contour_w07}.
The reduced chi-square of the velocity dispersion fit of each dSph lies in
the range $0.61 \-- 1.66$.
Note that the estimates of $m_{\psi}$ are, in general, lower than those obtained
from W09, which arises from the generally higher velocity dispersions in W07.
We address this discrepancy in more detail in \sref{subsubsec:VelDisp}.
By multiplying the one-dimensional posterior
distributions of the seven dSphs in W07, we get $1\sigma\,(2\sigma)$
confidence intervals of $m_{\psi}=1.18_{-0.13(-0.24)}^{+0.14(+0.28)}\times10^{-22}\eV$
(see Appendix \ref{sec:Joint analysis} for the results of fitting
all dSphs in W07 simultaneously).
This estimate is marginally consistent with the lower limit from W09. The
corresponding reduced chi-square of particle mass is $\chi_{\rm{red}}^2=0.92$. The estimated means,
$1\sigma$, and $2\sigma$ confidence intervals of $r_c$, $m_{\psi}$, and $\beta$
for each dSph are also listed in \tref{table:fit}.

\section{Discussion}
\label{sec:Discussion}

\subsection{Model Uncertainties}
\label{subsec:ModelUncertainty}

In order to consolidate the results of our Jeans analysis, we test how
sensitive $m_{\psi}$ is to different models adopted in the MCMC calculations,
including different observational data sets of the velocity dispersion profiles,
different stellar density models, and a different orbital anisotropy prior distribution.
Results are summarized in \tref{table:discussion}.

\subsubsection{Velocity Dispersion}
\label{subsubsec:VelDisp}

\begin{figure}
\centering
\includegraphics[width=\columnwidth]{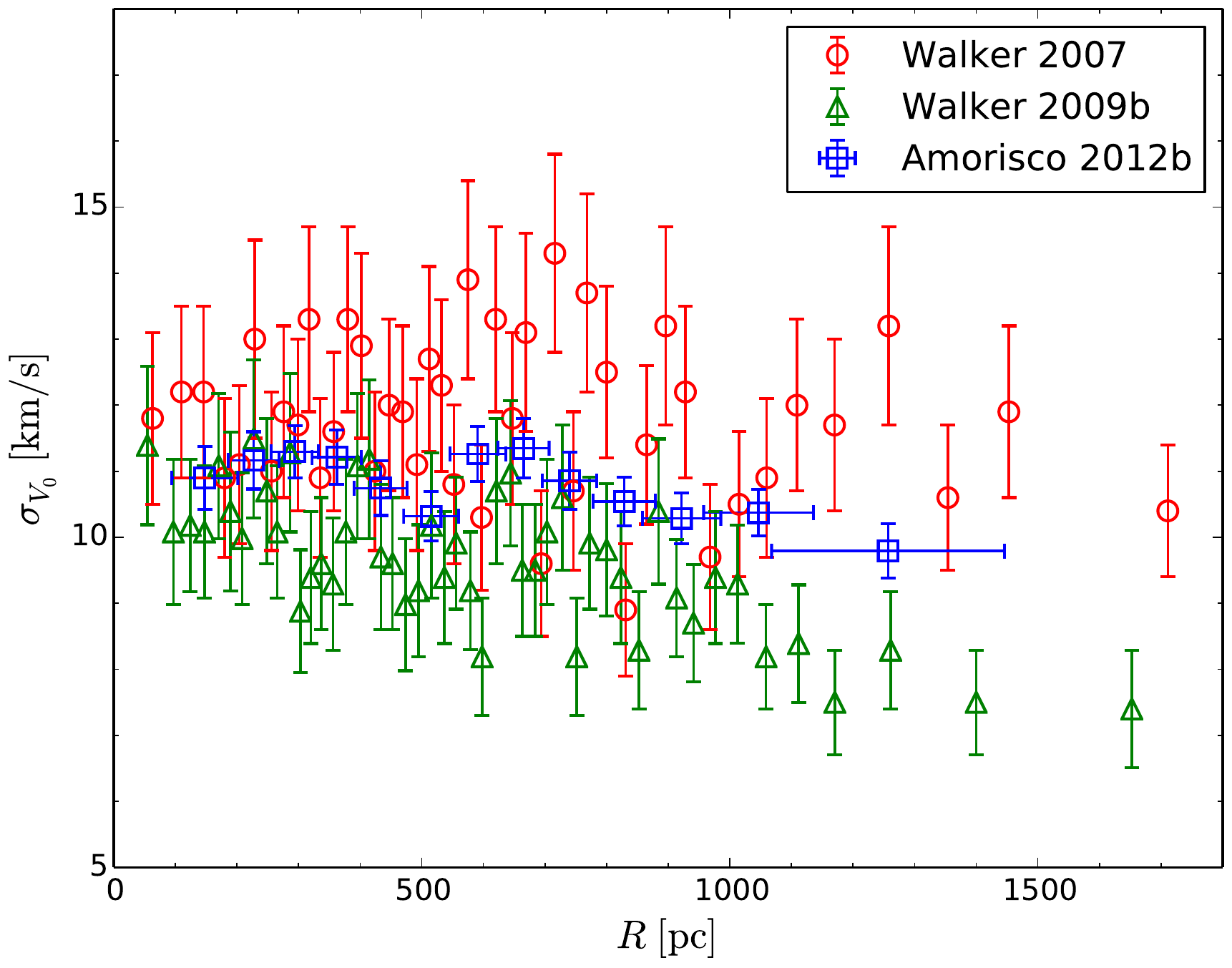}
\caption{
Comparison of the projected velocity dispersion profiles of Fornax given by
\citet{Walker2007} (circles), \citet{Walker2009b} (triangles), and
\citet{Amorisco&Evans2012b} (squares).
}
\label{fig:fornax_dispersion}
\end{figure}

As shown in \tref{table:fit},
the estimates of $m_{\psi}$ based on the velocity dispersion profiles of W07
are, in general, lower than those obtained from W09. It results from the
generally higher velocity dispersions in W07. This difference is most
significant in Fornax, where only about half of the data points in W07 and W09
overlap within the $1\sigma$ range (see Fig. \ref{fig:fornax_dispersion}).
For comparison, we also show the velocity dispersion profile of
\citet{Amorisco&Evans2012b}, which lies in between W07 and W09.
The inverse-variance-weighted means of the velocity dispersions of Fornax in
W07 and W09 are $\bar{\sigma}_{V_0}=11.6\pm 0.2$ and
$\bar{\sigma}_{V_0}=9.4\pm 0.1$, respectively.

The lower velocity dispersion in W09 mainly results from a more restrictive
selection of member stars, which only includes stars with a membership
probability greater than $95\%$ \citep{Walker2009b, Walker2009c}. The discarded
stars can have velocity $3\sigma$ away from the mean velocity (see Figure 1 and
Table 1 in \citet{Walker2009c}), and thus discarding them would lower the velocity
dispersion noticeably.
Also, the new samples of Fornax in W09 lead to a gently declining velocity dispersion profile
at large projected radii $(\geq 1 \kpc)$ as compared with W07.
Since Fornax provides the most stringent constraints on $m_{\psi}$, any
difference in the observed velocity dispersion of Fornax would directly
effect the estimate of $m_{\psi}$.

\begin{figure}
\centering
\includegraphics[width=\columnwidth]{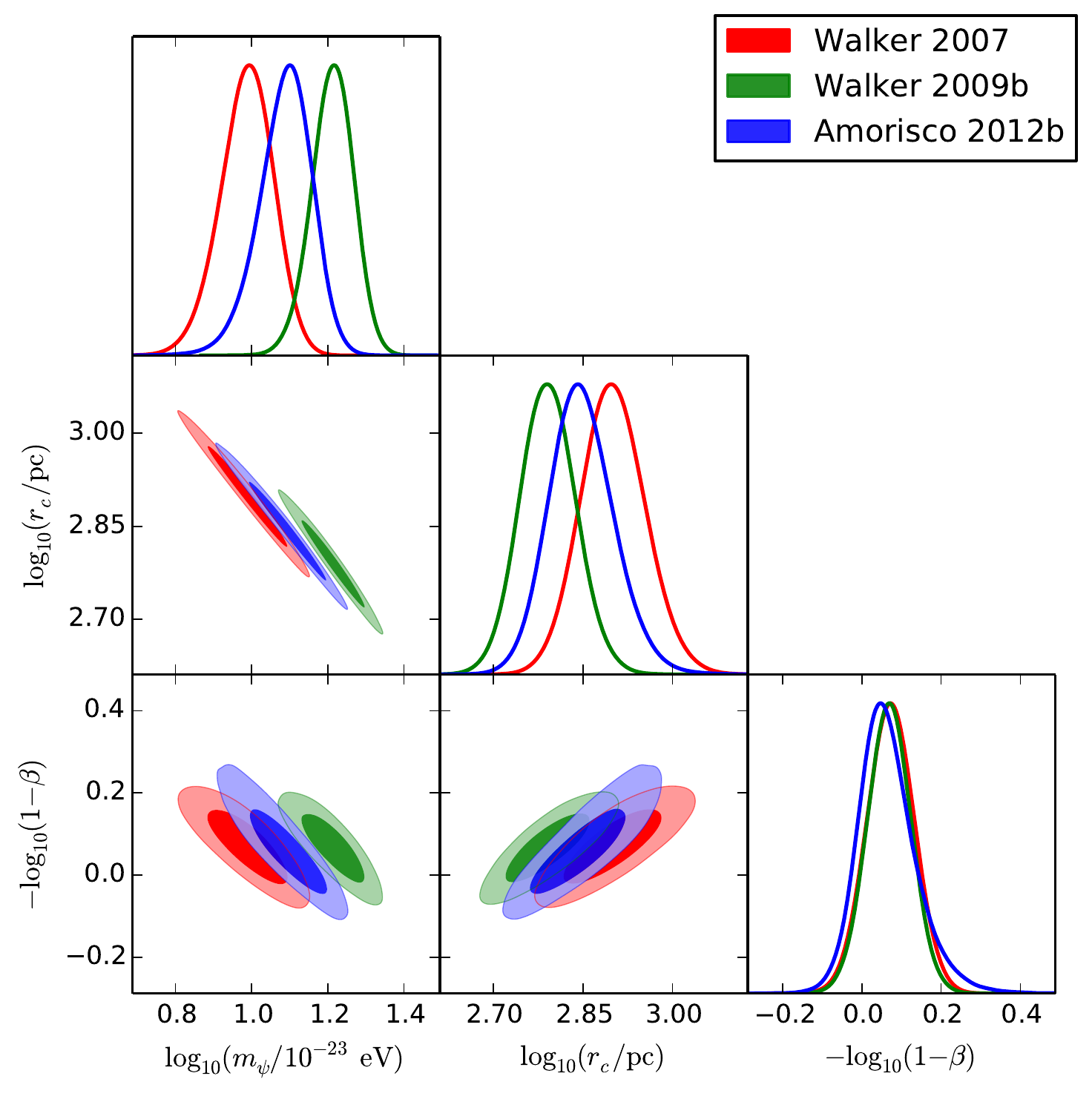}
\caption{
Posterior distributions of $m_{\psi}$, $r_{c}$, and $\beta$ in our MCMC
analysis for the three different velocity dispersion data sets
of Fornax shown in \fref{fig:fornax_dispersion}. Solid curves in the diagonal
panels show the one-dimensional marginalized distributions. Filled contours
in the corner panels show the $1\sigma$ and $2\sigma$ confidence regions.
}
\label{fig:fornax_fit}
\end{figure}

\fref{fig:fornax_fit} shows a comparison of the one-dimensional and
two-dimensional posterior distributions of $m_{\psi}$, $r_{c}$, and $\beta$
from the three different velocity dispersion profiles of Fornax in
\fref{fig:fornax_dispersion}. It clearly shows that the estimate of $m_{\psi}$
is sensitive to the adopted velocity dispersion profile, where a lower and steeper
profile leads to a higher estimate of $m_{\psi}$. The data sets of W07 and W09
thus likely bracket the uncertainty of the estimates of $m_{\psi}$, which
lies in the range $m_{\psi} \sim 1 \-- 2\times10^{-22}\eV$. We emphasize that
this value is in good agreement with other independent estimates from the stellar
subpopulations in dSphs \citep{Schive2014a, Marsh&Pop2015}, the high-redshift
luminosity functions \citep{Bozek2015,Schive2016,Corasaniti2016}, and the
Thomson optical depth to the cosmic microwave background \citep{Bozek2015,Schive2016}.

We also notice that the estimate of $m_{\psi}$ from Sextans in W09 is
higher than those obtained from other dSphs in both W07 and W09.
For example, by using the data of Sextans in W07, we find a $1\sigma$
confidence interval of $\log_{10}[m_{\psi}/10^{-23}\rm{eV}]= 1.33_{-0.13}^{+0.22}$,
significantly lower than the value determined using W09 but more consistent
with other dSphs (see \tref{table:fit}). It is mainly because the inverse-variance-weighted mean
of the velocity dispersion of W07 is $\bar{\sigma}_{V_0}=7.1\pm 0.3$,
apparently higher than that of W09, $\bar{\sigma}_{V_0}=6.1\pm 0.3$.

To look into this discrepancy, we further apply the same MCMC algorithm to
the more recent Very Large Telescope (VLT) observation of Sextans
\citep{Battaglia2011}, which is in better agreement with W07 than W09.
We find $r_{c} \sim 1.5-3.5\kpc$, consistent with \citet{Battaglia2011} using a
pseudo-isothermal model for the dark matter density profile. However, we get
$\log_{10}[m_{\psi}/10^{-23}\rm{eV}]= 0.51_{-0.44}^{+0.23}$ in a $1\sigma$
confidence interval, apparently lower than the estimates from both W07 and W09.
The inconsistency between $m_{\psi}$ determined from the three different
observations therefore suggest that the current data of Sextans provide a poor
constraint on $m_{\psi}$.

One caveat about the VLT data is that they extend to as far as $\sim 2.5\kpc$,
which lies beyond $3\,r_c$ for the $r_c$ estimated from W07.
It thus seems reasonable to consider an NFW halo outside the central soliton.
However, the NFW halo introduces two additional free parameters, making the
Jeans analysis infeasible due to the very few data points.

There is evidence of tidal stripping in the outermost parts of both Carina
and Leo I \citep{Munoz2006,Mateo2008,Battaglia2012}. Therefore, for these two dSphs we also
apply Jeans analysis discarding the outermost one data point in the velocity
dispersions. We obtain $1\sigma$ confidence intervals of
$\log_{10}[m_{\psi}/10^{-23}\rm{eV}]=1.41_{-0.18}^{+0.44}$ for Carina and
$\log_{10}[m_{\psi}/10^{-23}\rm{eV}]=1.19_{-0.20}^{+0.32}$ for Leo I,
both consistent with \tref{table:fit}. However, we notice the relatively
larger ranges in the revised estimates of $m_{\psi}$, which arise from
the $m_{\psi}\--r_c$ degeneracy.
For example, for Carina the outermost data point has a projected radius of
$R\sim 870\pc$, well beyond the estimated mean core radius, $r_c = 570\pc$.
For comparison, the second outermost data point has a projected radius of
$R\sim 530\pc$, comparable to the core radius. Therefore, the outermost data
point is important for constraining the dark matter mass profile and breaking
the $m_{\psi}\--r_c$ degeneracy.

\subsubsection{Stellar Density}
\label{subsubsec:StellarDens}

It is important to investigate the impact of different stellar density models.
First, in addition to having $R_h$ randomly sampled from a Gaussian distribution with
a given mean and variance, we also experiment with fixing $R_h$ to the mean
value and validate that the estimates of $m_{\psi}$ are largely unchanged in
all dSphs.
Second, we adopt a generalized Plummer model \citep{Mashchenko2015},
$I(R)=L_{0} \left[ 1 + (R/b)^2\right]^{-(\alpha-1)/2}$,
where $L_{0}$ is the central luminosity, $b$ is the core radius, and $\alpha$
is an integer which must be greater than three for a finite total stellar mass.
The standard Plummer model corresponds to $\alpha=5$.
This model has been shown to fit well with recent observations of
Fornax \citep{Coleman2005}, Leo I \citep{Smolcic2007}, and
Draco \citep{Odenkirchen2001} even at the tidal radius.
The corresponding estimates of particle mass are
$\log_{10}[m_{\psi}/10^{-23}\rm{eV}]= 1.15_{-0.06}^{+0.07}$ for Fornax,
$\log_{10}[m_{\psi}/10^{-23}\rm{eV}]= 0.92_{-0.21}^{+0.42}$ for Leo I, and
$\log_{10}[m_{\psi}/10^{-23}\rm{eV}]= 0.85_{-0.19}^{+0.34}$ for Draco,
in good agreement with \tref{table:fit}. A relatively larger difference is found
in Leo I. It is because the data of \citep{Smolcic2007}, when fitted with
a King profile, suggests a $50\%$ larger King core radius but a similar tidal
radius compared to the data of \citet{Irwin1995}.
Finally, for Sextans, we also apply an exponential profile \citep{Irwin1995}.
It leads to $\log_{10}[m_{\psi}/10^{-23}\rm{eV}]= 1.64_{-0.14}^{+0.48}$,
consistent with \tref{table:fit}.

\subsubsection{Orbital Anisotropy}
\label{subsubsec:Anisotropy}
For the orbital anisotropy, we investigate this issue with a different uniform prior,
$\eta=(\bar{v_r^2}-\bar{v_\theta^2})/(\bar{v_r^2}+\bar{v_\theta^2})$
\citep{Mashchenko2006}. This form has the advantage of being symmetric for
tangential and radial anisotropies, where $\eta=-1$ for pure tangential orbit,
$\eta=0$ for isotropic orbit, and $\eta=1$ for pure radial orbit. We validate
that the resulting $m_{\psi}$ is very consistent with \tref{table:fit}.

One caveat in Jeans analysis is that the choice of functional form for
stellar anisotropy may affect the estimate of enclosed mass due to the
mass-anisotropy degeneracy \citep[e.g.,][]{Gonzles2016}. Using more flexible dynamical modeling
methods that extract more information from the line-of-sight velocity or
stellar distributions, such as modeling higher moments \citep{Lokas2002},
phase-space analyses \citep{Kleyna2002,Amorisco&Evans2012a}, and Schwarzschild's orbit
superposition methods \citep{Jardel2012,Breddels2013}, can help
reduce the degeneracy and provide a more robust constraint on $m_{\psi}$.
We plan to address this issue in the future.

\subsection{Soliton + NFW model}
\label{subsec:SolitonNFW}

To justify the assumption that all stars are located within the
central soliton core, we investigate the case where the soliton core
connects to an NFW halo at a larger radius. The overall density profile,
as shown by \citet{Schive2014a,Schive2014b,Marsh&Pop2015}, can be modeled as
\begin{equation}
\rho(r)=\Theta (r_\epsilon -r)\rho_{\rm{soliton}}(r) +
        \Theta (r-r_\epsilon)\rho_{\rm{NFW}},
\label{eq:soliton_plu_NFW}
\end{equation}
where
\begin{equation}
\rho_{\rm{NFW}}(r)=\frac{\rho_0}{(\frac{r}{r_s})(1+\frac{r}{r_s})^{2}}
\label{eq:NFW}
\end{equation}
is the NFW profile, $\rho_{\rm{soliton}}(r)$ is given in \eref{eq:SolitonFit},
$\Theta(r)$ is a step function, and $r_\epsilon$ is the transition radius
between soliton and NFW halo given by
$\rho_{\rm{NFW}}(r_\epsilon) = \rho_{\rm{soliton}}(r_\epsilon) = \epsilon\rho_{\rm{soliton}}(0)$.
Since $\rho_0$ can be uniquely determined from a given $r_\epsilon$ and $r_s$,
this model introduces two additional free parameters,
$\left \{ \log_{10}\epsilon, \log_{10}[r_s/\rm{pc}]\right \}$,
which we model with uniform priors over the ranges
$-5 \leq \log_{10}\epsilon \leq \log_{10}0.5$ and
$1 \leq \log_{10}[r_s/\rm{pc}] \leq 4$.

Note that the soliton core mass $(M_c)$ has been found to correlate with halo
virial mass $(M_h)$ in the mass range $\sim 10^8 \-- 5\times10^{11}\Msun$
\citep{Schive2014b} as
$M_c = 0.25(M_h/M_{\rm{min},0})^{1/3} M_{\rm{min},0}$,
where $M_{\rm{min},0} \sim 4.4\times10^7 (m_{\psi}/10^{-22}\eV)^{-3/2}\Msun$
is the predicted minimum $\psiDM$ halo mass at the present day.
Using this relation can, in principle, eliminate one free
parameter. However, it is unclear whether this relation,
which is determined from isolated galaxies, can be applied to satellite dSph
galaxies underwent complex evolution history. In order to make our analysis
more robust, we thus do not take this core-halo mass relation into account
when conducting Jeans analysis.

We apply this soliton + NFW model to Fornax since it provides the strongest constraint on
$m_{\psi}$ and has around half of the observation points lying outside
the core radius determined from the soliton-only model. To properly constrain
the central soliton density profile that has two free parameters, we set the
lower limit of $r_c$ to the position of the second innermost observation
point, which is $\sim 100\pc$. This is a very weak constraint since the core
size of Fornax is typically found to be $\sim 1\kpc$
\citep[e.g.,][]{Amorisco2013, Walker2011}. We obtain
$\log_{10}[m_{\psi}/10^{-23}\rm{eV}]=1.23_{-0.07(-0.14)}^{+0.05(+0.13)}$,
and $\log_{10}[r_c/\rm{pc}]=2.78_{-0.05(-0.11)}^{+0.06(+0.11)}$
(also listed in \tref{table:discussion}),
consistent with the soliton-only model (see \tref{table:fit}). The
transition radius between soliton and NFW halo is found to be greater
than $\sim 2.5\,r_c$, where the soliton density has dropped by a factor of 30.
The scale radius $r_s$ is unconstrained, which is expected since most stars still reside within
the transition radius. The slightly larger $m_{\psi}$ and smaller $r_c$ arise
from the additional gravity support from the external NFW halo for stars
outside the transition radius.

\subsection{Stellar Subpopulations}
\label{subsec:subpopulation}

\begin{figure}
\centering
\includegraphics[width=\columnwidth]{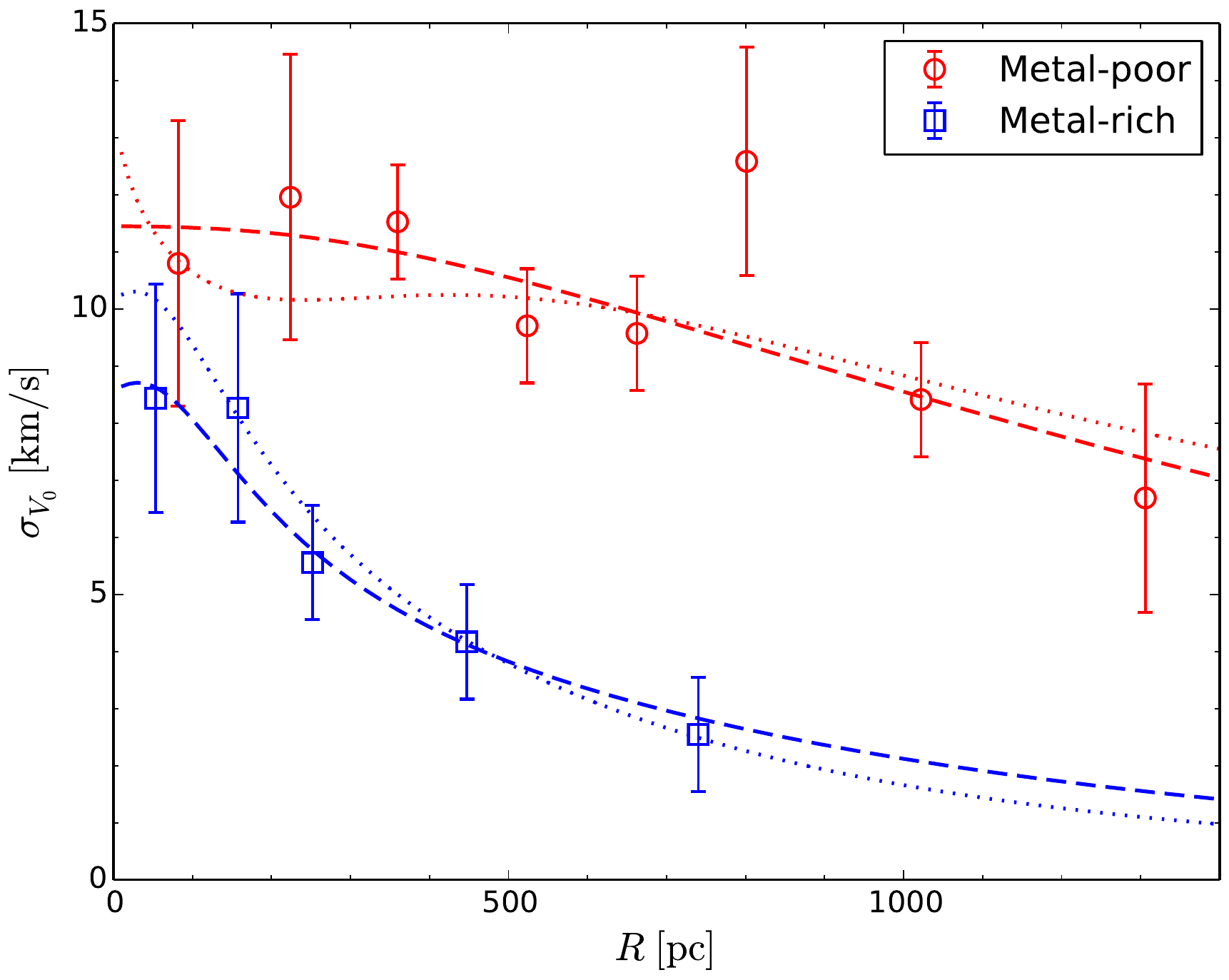}
\caption{
Projected velocity dispersion profiles of the two stellar subpopulations
in Sculptor. Error bars show the observational $1\sigma$ uncertainties
\citep{Battaglia2008}. Dashed (dotted) lines show the best-fit $\psiDM$
model in the hypothesis of OM (constant) anisotropy profile on the MP subpopulation
and OM anisotropy profile on the MR subpopulation. 
}
\label{fig:Sculptor_dispersion_2population}
\end{figure}

Several dSphs have been found to exhibit more than one stellar
subpopulations, each with different metallicity and kinematics. Modeling
different subpopulations separately in the same gravitational potential
can further break the mass-anisotropy degeneracy in the Jeans analysis \citep{Battaglia2008}. 
We model the stellar density and velocity dispersion profiles of the
metal-rich (MR) and metal-poor (MP) subpopulations in Sculptor \citep{Battaglia2008}.
The MR subpopulation is known to be better described by radially biased orbits
in the outer region due to its rapidly declining velocity dispersion profile
\citep{Battaglia2008, Strigari2014}.
Therefore, we follow \citet{Battaglia2008} and adopt the Osipkov-Merritt
\citep[OM,][]{Osipkov1979,Merritt1985} anisotropy profile on the MR
subpopulation, and use either constant or OM anisotropy on the MP
subpopulation. The OM anisotropy profile is given by
$\beta(r)=r^{2}/(r^{2}+r_{a}^{2})$, where $r_a$ is the anisotropy radius with
$\beta \rightarrow 0$ for $r \ll r_a$ and $\beta \rightarrow 1$ for $r \gg r_a$.
We discard the last observation data point of the
MP subpopulation because of its large uncertainty.
The $1\sigma$ ranges for constant and OM anisotropy on the MP subpopulation are
$\log_{10}[m_{\psi}/10^{-23}\rm{eV}]=1.21_{-0.15}^{+0.17}$ and
$\log_{10}[m_{\psi}/10^{-23}\rm{eV}]=1.08_{-0.22}^{+0.28}$, respectively
(also listed in \tref{table:discussion}).
Both are in good agreement with \tref{table:fit}. 
The empirical and the best-fit velocity dispersion profiles in our MCMC chains are 
shown in \fref{fig:Sculptor_dispersion_2population}, which have chi-squares of
$\chi^2=6.8$ (with $\chi_{\rm{MP}}^{2}=5.4,\,\chi_{\rm{MR}}^{2}=1.4$) and
$\chi^2=4.3$ (with $\chi_{\rm{MP}}^{2}=3.8,\,\chi_{\rm{MR}}^{2}=0.5$)
for constant and OM anisotropy on the MP subpopulation, respectively.

Fornax also has three distinct stellar subpopulations
\citep{Amorisco&Evans2012c}. The intrinsic rotation detected in the MP and
intermediate-metallicity (IM) subpopulations is
${\Omega}_{int}\sim 1-2\kms\kpc^{-1}$, which
is negligible compared with velocity dispersion. For example, at the
outermost kinematic data point ($\sim 1\kpc$), the ratio between rotation
velocity and velocity dispersion is only $\sim 0.3$ \citep{Amorisco&Evans2012c}. 
Therefore, we neglect rotation and apply Jeans analysis to the three subpopulations using the
stellar density and velocity dispersion profiles in \citet{Amorisco2013}.

Note that \citet{Amorisco2013} applied an empirical relation between
$M(<1.67R_h)$, $R_h$, and $\bar{\sigma}_{V_0}$ to the three subpopulations
of Fornax to estimate the total mass profile (contours in \fref{fig:fornax_fit_3population}).
This empirical mass estimator
has been derived to describe a wide family of models based on the Michie-King
phase-space distribution function \citep{Amorisco&Evans2012a}, where
velocity distribution is isotropic in the center and nearly radial in the outer
region, similar to the OM anisotropy model. Therefore, in order to compare
with their result, we adopt OM anisotropy profiles on all three stellar
subpopulations. In addition,
since the kinematic data have observational uncertainties in both $\sigma_{V_0}(R_i)$
and $R_i$, we follow \citet{Ma2013} and convert the uncertainty associated
with $R_i$ into an effective variance in $\sigma_{V_0}(R_i)$ as
$\rm{Var}[\sigma_{V_0}(R_i)]+{\sigma}'_p(R_i)\rm{Var}[R_i]$, where
${\sigma}'_p(R_i)$ is the derivative of the estimated velocity dispersions.

\begin{figure}
\centering
\includegraphics[width=\columnwidth]{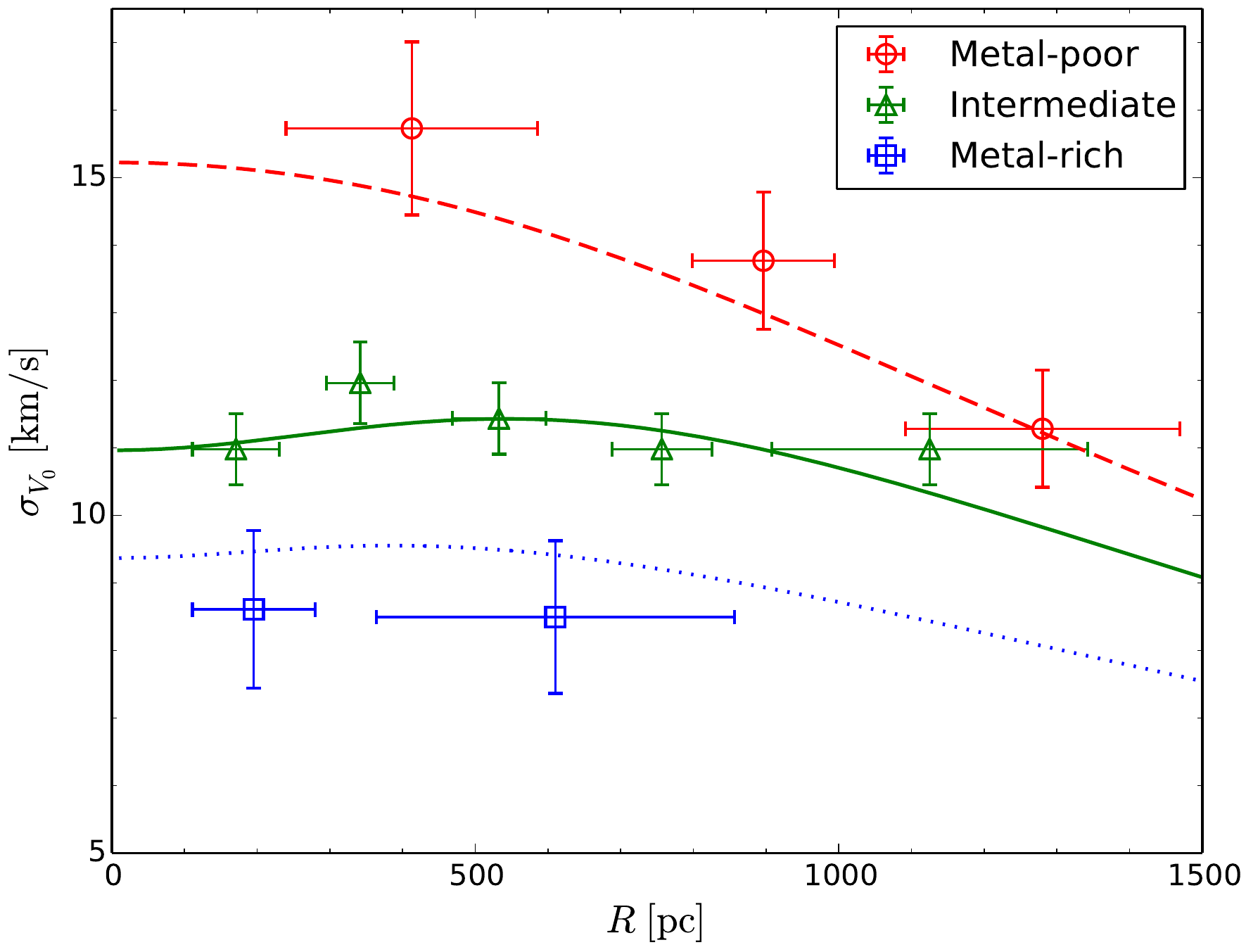}
\caption{
Projected velocity dispersion profiles of the three stellar subpopulations
in Fornax. Error bars show the observational $1\sigma$ uncertainties
\citep{Amorisco2013}, and lines show the best-fit profiles
within the $1\sigma$ ranges of our estimated $m_{\psi}$ and $r_c$ based on the soliton 
density profile in $\psiDM$.
}
\label{fig:fornax_dispersion_3population}
\end{figure}

\begin{figure}
\centering
\includegraphics[width=\columnwidth]{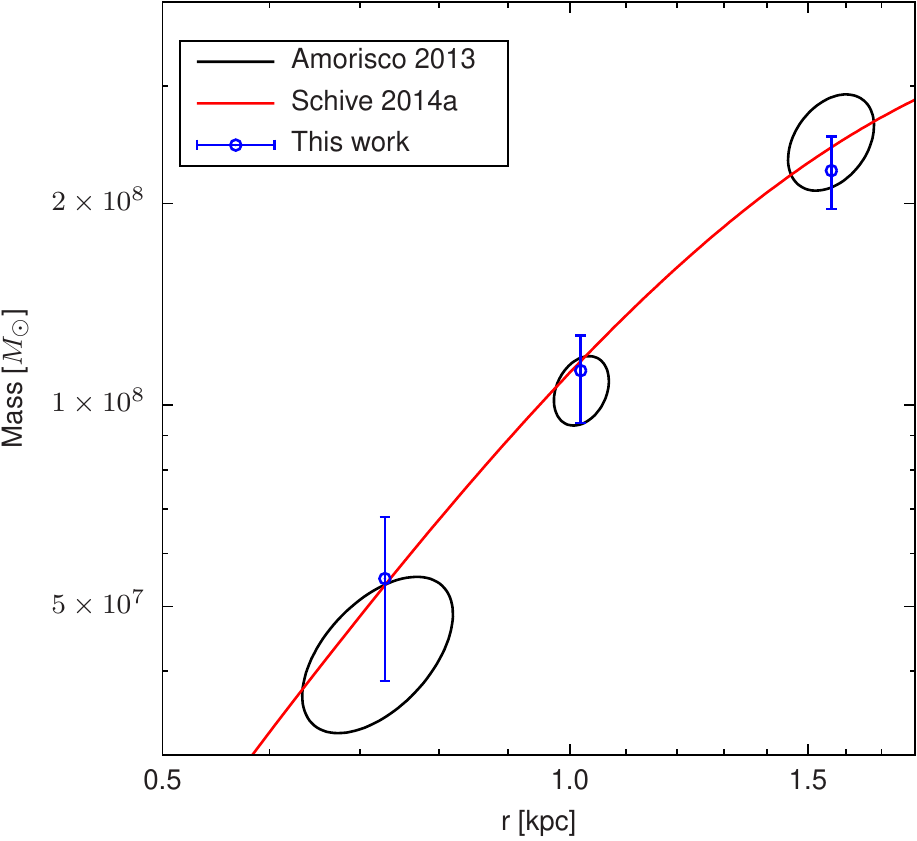}
\caption{
Total enclosed mass estimated from the three stellar subpopulations
shown in \fref{fig:fornax_dispersion_3population}. Error bars represent
the $1\sigma$ confidence intervals estimated in this work.
For comparison, solid contours show the $1\sigma$ estimate of \citet{Amorisco2013},
and solid line shows the best-fit soliton model using only the
intermediate-metallicity subpopulation \citet{Schive2014a}. These results
are in good agreement with each other.
}
\label{fig:fornax_fit_3population}
\end{figure}

We obtain $1\sigma$ confidence intervals of
$\log_{10}[m_{\psi}/10^{-23}\rm{eV}]= 0.91_{-0.09}^{+0.26}$ and
$\log_{10}[r_c/\rm{pc}]=2.96_{-0.19}^{+0.07}$.
\fref{fig:fornax_dispersion_3population} shows the empirical velocity
dispersions of the three subpopulations and our estimated velocity
dispersions, which correspond to the maximum likelihood point within the
$1\sigma$ range. \fref{fig:fornax_fit_3population} shows the corresponding $1\sigma$ 
total enclosed mass, $M(<1.67R_h)$, where $R_h$ is the half-light radius of
each subpopulation. This result is in good agreement with the $1\sigma$
estimate of \citet[][contours in \fref{fig:fornax_fit_3population}]{Amorisco2013}
It is also consistent with the estimate of
\citet[][solid line in the figure]{Schive2014a} using only the IM
subpopulation, which gives $m_{\psi}=8.1_{-1.7}^{+1.6}\times10^{-23}\eV$
and $r_c=0.92_{-0.11}^{+0.15}\kpc$

Note that we find another local maximum of likelihood function around 
$\log_{10}[m_{\psi}/10^{-23}\rm{eV}]=0.59_{-0.15}^{+0.18}$ and
$\log_{10}[r_c/\rm{pc}]= 3.18_{-0.11}^{+0.10}$,
with anisotropy radii of the IM and MP subpopulations close to their $R_h$.
However, this local peak only covers a small volume in the five-dimensional
posterior distributions and is negligible in the one-dimensional marginalized
distributions of $m_{\psi}$ and $r_c$.
Also note that the MR subpopulation, which has the lowest velocity dispersion,
has only two observation points and thus gives
a relatively weak constraint compared with the IM and MP subpopulations.
Accordingly, the estimate of $m_{\psi}$ is more consistent with that using W07,
which has a higher velocity dispersion closer to IM and MP than to MR.

\section{Conclusion}
\label{sec:Conclusion}
Wave dark matter ($\psiDM$), characterized by a single parameter, the dark
matter particle mass $m_{\psi}$, predicts a central soliton core in every galaxy.
In this work, we have applied the Jeans equation to the empirical velocity
dispersion profiles of eight classical dSphs in the Milky Way
so as to constrain $m_{\psi}$. We find combined $1\sigma (2\sigma)$
confidence intervals of
$m_{\psi}=1.18_{-0.13(-0.24)}^{+0.14(+0.28)}\times10^{-22}\eV$ and 
$m_{\psi}=1.79_{-0.17(-0.33)}^{+0.17(+0.35)}\times10^{-22}\eV$ using the
observational data sets of \citet{Walker2007} and \citet{Walker2009b},
respectively. The discrepancy of W07 and W09 suggests 
that a more elaborate star membership determination for calculating velocity dispersion
will improve the constraint on $m_{\psi}$ further.

This combined constraint of $m_{\psi}$ is dominated by Fornax but is consistent
with the estimate from individual dSphs. It is also
in good agreement with other independent constraints from, 
for instance, the stellar subpopulations in dSphs
\citep{Schive2014a,Marsh&Pop2015}, the high-redshift luminosity functions \citep{Bozek2015,Schive2016}, and
the Thomson optical depth to CMB \citep{Bozek2015,Schive2016},
which all suggest $m_{\psi} \sim 10^{-22}\eV$.

To consolidate the results, we have investigated a variety of models in the
MCMC calculations, including different velocity dispersion data, stellar
density profiles, and orbital anisotropy priors. We have also extended the
soliton-only model to account for an NFW halo at a larger radius, and further
considered distinct stellar subpopulations
in both Sculptor and Fornax. It is demonstrated that these factors have only
minor effect on the estimate of $m_{\psi}$.

Finally, we emphasize that the existence of large cores in dSphs is still
under debate \citep[e.g.,][]{Strigari2010}. The methodology of our study is to
assume a soliton core profile, \eref{eq:SolitonFit}, and ascertain whether the
resulting constraint on $m_{\psi}$ is consistent with other independent constraints
mentioned above. In other words, we focus on validating the self-consistency of
$\psiDM$, but not on falsifying the NFW profile or the CDM model.
In principle, the latter can be addressed by extending the soliton + NFW model
to allow for a much smaller soliton component
and we plan to explore it in the future.

\section{Acknowledgement}
This work is supported in part by the National Science Council of Taiwan
under the grant MOST 103-2112-M-002-020-MY3.

\bibliographystyle{mnras}
\bibliography{Reference} 

\appendix
\section{Soliton Mass Profile}
\label{sec:SolitonMassProfile}

The soliton mass profile can be explicitly derived from \eref{eq:SolitonFit} as
\begin{multline}
M_{\rm{soliton}}(r) = \frac{4.2\times10^9\Msun}{(m_{\psi}/10^{-23}\eV)^{2} (r_c/\pc)} \frac{1}{(a^{2} + 1)^{7}} \\
                      (3465 a^{13} + 23100 a^{11} + 65373 a^{9} + 101376 a^{7} + 92323 a^{5} \\
                      + 48580 a^{3} - 3465 a + 3465 (a^{2} + 1)^{7} \arctan{(a)}),
\label{eq:soliton_mass_profile}
\end{multline}
where $a=(2^{1/8}-1)^{1/2} (r/r_c)$.

\section{Joint analysis}
\label{sec:Joint analysis}

To consolidate our estimates of $m_{\psi}$, here we perform the joint analysis
by fitting all dSphs \emph{simultaneously}.
Assuming all dSphs are independent data sets, the joint likelihood function is the product
of the likelihood functions of individual dSphs.
We treat the particle mass $m_{\psi}$ as a global parameter shared by all dSphs,
while allow $r_c$, $\beta$ to vary in different dSphs.
To account for the uncertainty in $R_h$, we adopt Gaussian priors with means and
standard deviations equal to the observations, and then integrate it out when calculating
the posterior distributions of other model parameters.

\begin{figure}
\centering
\includegraphics[width=\columnwidth]{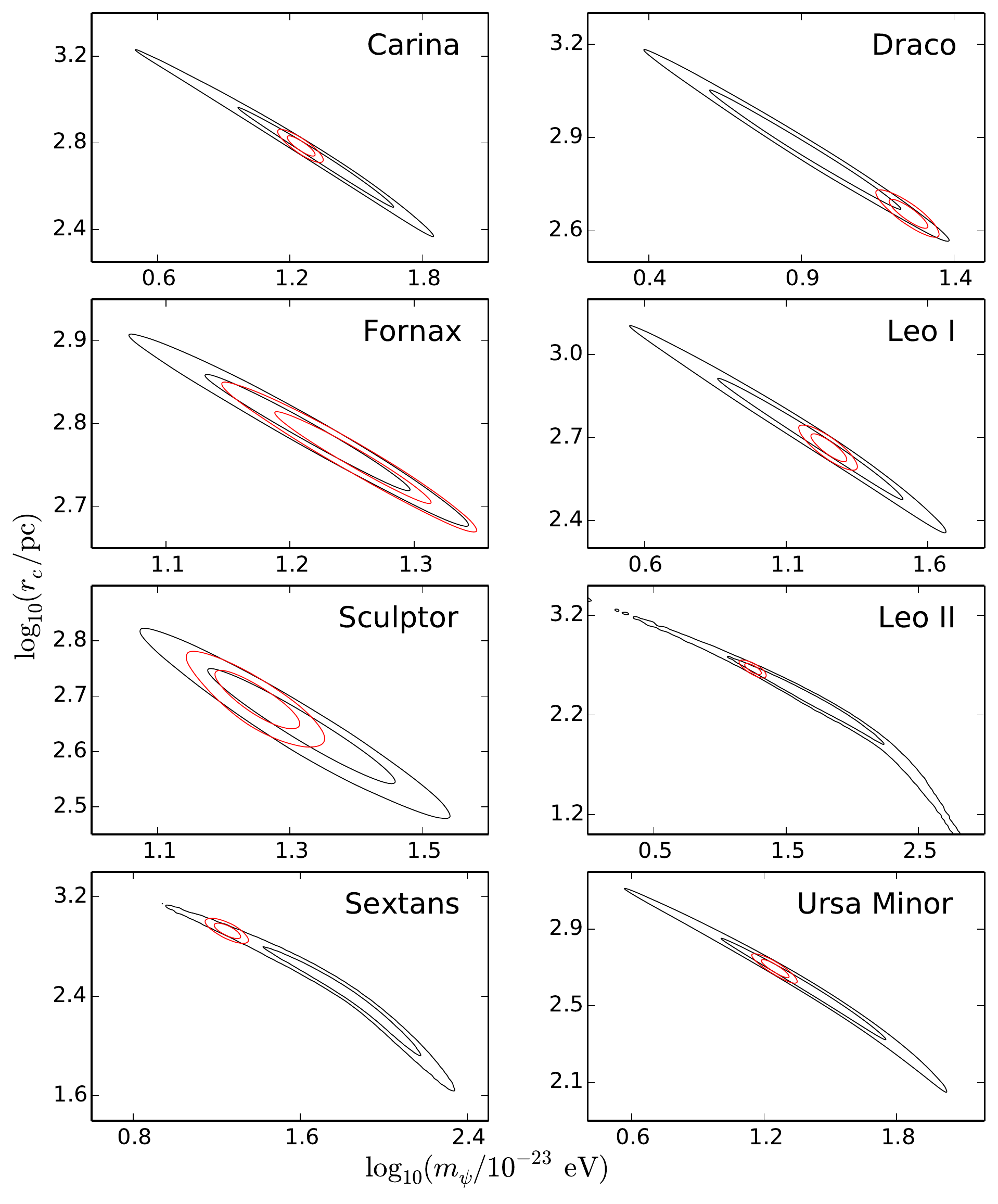}
\caption{
Posterior distributions of $m_{\psi}$ and $r_{c}$ obtained from the joint analysis (red lines),
overplotted with \fref{fig:mb_rc_contour_w09} obtained from analyzing individual dSphs separately (black lines).
Contours show the $1\sigma$ and $2\sigma$ confidence regions.
}
\label{fig:mb_rc_combine09}
\end{figure}

\begin{figure}
\centering
\includegraphics[width=\columnwidth]{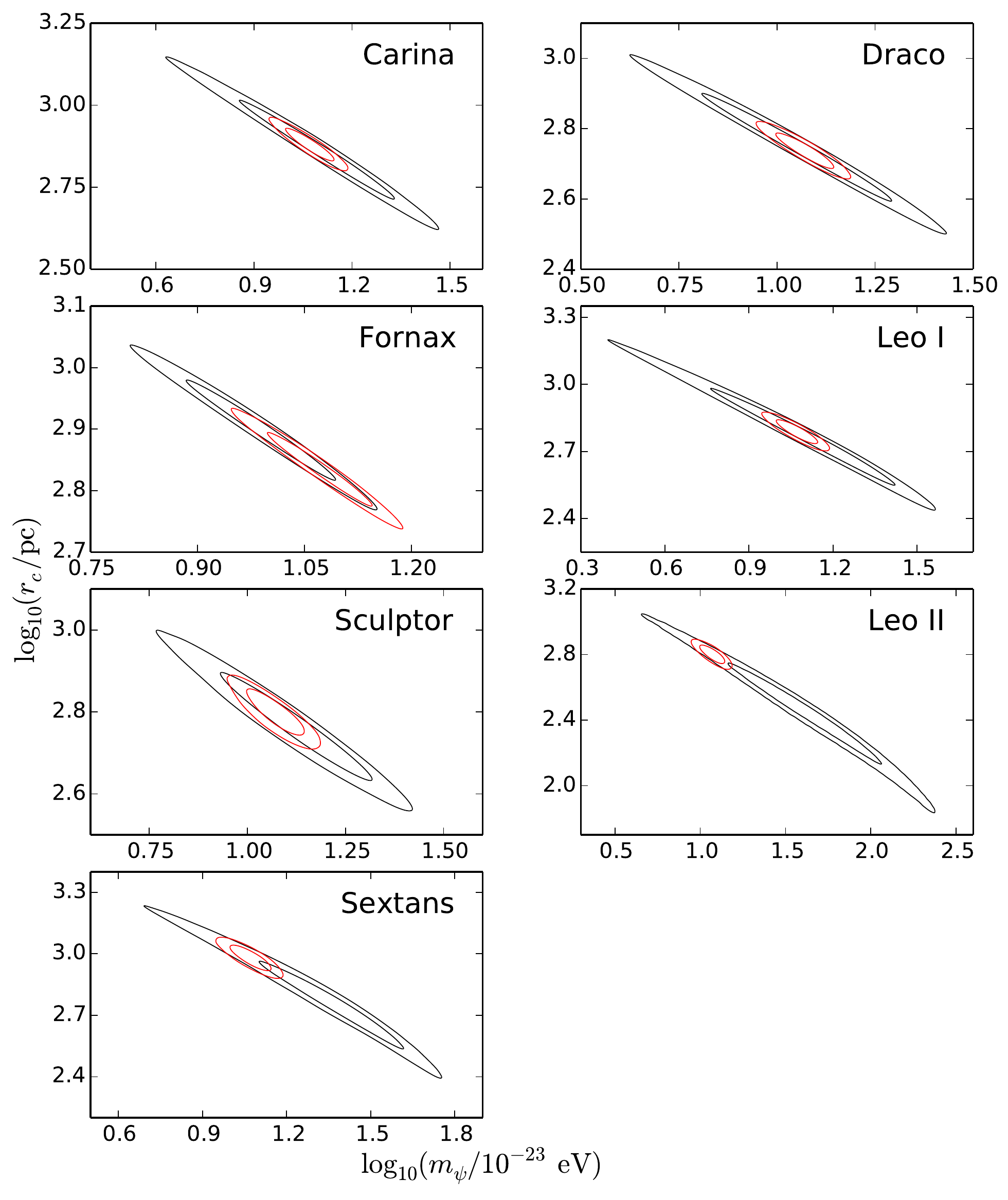}
\caption{
Same as \fref{fig:mb_rc_combine09} but for the observational data set of \citet{Walker2007}.
}
\label{fig:mb_rc_combine07}
\end{figure}

Figures \ref{fig:mb_rc_combine09} and \ref{fig:mb_rc_combine07} show the 
$m_{\psi}$-$r_{c}$ confidence contours from the joint analysis, overplotted with the previous results 
shown in Figures \ref{fig:mb_rc_contour_w09} and \ref{fig:mb_rc_contour_w07} obtained from individual dSphs.
The $1\sigma\,(2\sigma)$ confidence intervals of $m_{\psi}$ is
$1.79_{-0.17(-0.33)}^{+0.17(+0.35)}\times10^{-22}\eV$ from the W09 data sets and
$1.18_{-0.13(-0.25)}^{+0.13(+0.28)}\times10^{-22}\eV$ from the W07 data sets.
These constraints are in good agreement with the estimates given in \sref{sec:Results}
where different dSphs are fitted separately.

\bsp
\label{lastpage}
\end{document}